# Are Happy Developers more Productive?

## The Correlation of Affective States of Software Developers and their self-assessed Productivity

Daniel Graziotin, Xiaofeng Wang, Pekka Abrahamsson

Free University of Bozen-Bolzano, Bolzano, Italy

{daniel.graziotin, xiaofeng.wang, pekka.abrahamsson}@unibz.it

**Abstract.** For decades now, it has been claimed that a way to improve software developers' productivity is to focus on people. Indeed, while human factors have been recognized in Software Engineering research, few empirical investigations have attempted to verify the claim. Development tasks are undertaken through cognitive processing abilities. Affective states – emotions, moods, and feelings - have an impact on work-related behaviors, cognitive processing activities, and the productivity of individuals. In this paper, we report an empirical study on the impact of affective states on software developers' performance while programming. Two affective states dimensions are positively correlated with self-assessed productivity. We demonstrate the value of applying psychometrics in Software Engineering studies and echo a call to valorize the human, individualized aspects of software developers. We introduce and validate a measurement instrument and a linear mixed-effects model to study the correlation of affective states and the productivity of software developers.

## 1 Introduction

For more than thirty years, it has been claimed that a way to improve software developers' productivity and software quality is to focus on people [4]. In more recent years, the advocates of Agile software development stress this to the point that "If the people on the project are good enough, they can use almost any process and accomplish their assignment. If they are not good enough, no process will repair their inadequacy – 'people trump process' is one way to say this." [6, p. 1].

Although research in productivity of software developers is well-established and rich in terms of proposals, little is still known on the productivity of individual programmers [30]. Nevertheless, there is an increasing awareness that human-related factors have an impact on software development productivity [29].

Arguably, human-related factors play an important role on software development as software artifacts are the result of intellectual activities. It is established but underestimated that software development is carried out through cognitive processing activities

[10, 16]. On the other hand, the role of affective states - i.e., emotions, moods, and feelings - in the workplace received significant attention in Management research and Psychology [1, 23, 36]. Affective states have an impact on cognitive activities of individuals [16]. Thus, it is necessary to understand how affective states play a role in software development.

In Software Engineering research, the inclination to study the human aspect of developers has also been translated to a call for empirical Software Engineering studies using psychometrics [9]. In particular, there is a call for research on the role of the affective states in Software Engineering [16, 31].

The research question that this study aims to answer is: how do the affective states related to a software development task in foci influence the self-assessed productivity of developers? To this end, we examine the variations of affective states and the self-assessed productivity of software developers while they are programming.

The main results of the study are two-fold. 1) The affective states of software developers are positively correlated with their self-assessed productivity. 2) The investigation produces evidence on the value of psychometrics in empirical Software Engineering studies.

This study offers an understanding, which is part of basic science in Software Engineering research rather than leading to direct, applicable results. However, with the added understanding on how affective states influence software developers, we are in a much better position to continue the pursuit for improving Software Engineering methods and practices.

The rest of this paper is structured as follows. Section 2 provides the background theory, the related work and the hypotheses of the study. Section 3 describes the research methodology, in order to be easily evaluated and replicated. Section 4 reports the outcomes of the experimental design execution. Section 5 contains the discussion of the obtained results and the limitations of the study. Section 6 concludes the paper with the theoretical implications of the study and suggestions for future research.

## 2 Related Work

### 2.1 Background Theory

Psychology and Cognitive Science have got a long history of studies in the field of psychometrics, affective states, and how individuals process information.

It is difficult to differentiate terms like affective states, emotions, and moods. Emotions have been defined as the states of mind that are raised by external stimuli and are directed toward the stimulus in the environment by which they are raised [25]. Moods have been defined as emotional states in which the individual feels good or bad, and either likes or dislikes what is happening around him/her [24]. However, there is still no clear agreement on the difference between emotion and mood. Many authors consider mood and emotion as interchangeable terms (e.g., [3], [8]). In this paper, we adopt the same stance and use the term affective states as a generic term to indicate emotions, moods, or feelings.

There are two main theories to categorize affective states. One theory, called the discrete approach, seeks a set of basic affective states that can be distinguished uniquely [25]. Examples include "interested", "excited", "upset", and "guilty". The other theory groups affective states in major dimensions, which allow clear distinction among them [28]. With this approach, affective states are characterized by their valence, arousal, and dominance. Valence (or pleasure) can be described as the attractiveness (or adverseness) of an event, object, or situation [20]. Arousal is the sensation of being mentally awake and reactive to stimuli, while dominance (or control, over-learning) is the sensation by which the individual's skills are higher than the challenge level for a task [7]. The dimensional approach is common in human-machine interaction and computational intelligence studies (e.g., [13, 33]). It is commonly adopted to assess affective states triggered by an immediate stimulus [5, 22]. Therefore, the dimensional approach is adopted in this study.

The measurement of affective states is usually achieved using questionnaires and surveys. One of the most used questionnaire for the dimensional approach is the Self-Assessment Manikin (SAM) [5, 18]. SAM is a non-verbal assessment method, based on pictures. SAM measures valence, arousal, and dominance associated with a person's affective reaction to a stimulus. A numeric value is assigned to each rating scale for each dimension. For a 5-point rating scale, a value of 5 for valence means "very high attractiveness and pleasure towards the stimulus". SAM is not uncommon in Computer Science research where the affective states towards a stimulus must be studied (e.g., [13]).

These scales and similar other psychometrics present issues when employed in within- and between-subjects analyses. There is not a stable and shared metric for assessing the affective states across persons. For example, a score of 1 in valence for a person may be equal to a score of 3 for another person. Nevertheless, it is sensible to assume a reasonable, stable metric within a person. To overcome this issue, the scores of each participant are converted to Z-scores (a.k.a. standard scores). An observation is expressed by how many standard deviations it is above or below the mean of the whole set of an individual's observations. In this way, the measurements between participants become dimensionless and comparable with each other [19].

The affective states of individuals have impact on work-related behaviors and capacities [1, 11, 15, 21]. Additionally, the positive-Psychology branch defines the mental status of flow as fully focused motivation, energized focus, full involvement, and success in the process of the activity [7]. The correlation with productivity seems straightforward. In fact, evidence has been found that happier employees are more productive [11, 23, 36].

## 2.2 Related Studies

The literature shows that the affective states have an impact on various cognitive activities of individuals and many of these activities are linked with software development.

Fisher and Noble [11] employ Experience Sampling Method [19] to study correlates of real-time performance and affective states while working. The study recruited different workers (e.g., child care worker, hairdresser, office worker); however none of them was reported to be a software developer. The measurement instrument was a questionnaire with 5 points Likert items. The paper analyzes self-assessed skills, task difficulty, affective states triggered by the working task, and task performance. It is not uncommon in Psychology to let participants self-evaluate themselves, as self-assessed performance is consistent to objective measurements of performance [21]. Among the results of the study, it is shown that there is a strong positive correlation between positive affective states and task performance while there is a strong negative correlation between negative affective states and task performance. This paper encourages further research about real-time performance and emotions.

Shaw [31] observes that, although the role of affective states in the workplace is a subject of studies in management theory, Information Technology research ignores the role of affective states on Information Technology professionals. The study shows that the affective states of a software developer may dramatically change during a period of 48 hours. However, the study is a work-in-progress paper and no continuation is known. Nevertheless, the study calls for research on the affective states of software developers.

Khan et al. [16] echo the previously reported call and provide links from Psychology and Cognitive Science studies to software development studies. The authors construct a theoretical two-dimensional mapping framework in two steps. In the first step, programming tasks are linked to cognitive tasks. For example, the process of constructing a program – e.g. modeling and implementation – is mapped to the cognitive tasks of memory, reasoning, and induction. In the second step, the same cognitive tasks are linked to affective states. The authors show a correlation with cognitive processing abilities and software development. Two empirical studies on affective states and software development are then reported, which relate a developer's debugging performance to the affective states. In the first study, affective states were induced to software developers, who were then asked to complete a quiz on software debugging. The second study was a controlled experiment. The participants were asked to write a trace on paper of the execution of algorithms implemented in Java. The results suggest that when valence is kept high and the arousal related to the task varies, there is a positive correlation with the debugging performance. This study recommends more research on the topic.

The body of knowledge suggests that affective states are positively correlated to the productivity of individuals. Therefore, the research hypotheses of this study are on positive correlations between real-time affective states and the immediate productivity of software developers. The following are the research hypotheses of this study.

— H1: The real-time valence affective state of software developers is positively correlated to their self-assessed productivity.
— H2: The real-time arousal affective state of software developers is positively correlated to their self-assessed productivity.
— H3: The real-time dominance affective state of software developers is positively correlated to their self-assessed productivity.

## 3 Research Methodology

The research methodology of this study is a series of repeated measurements in the context of multiple case studies on software developers, in which quantitative and qualitative data is gathered. In this section, we describe the design of the empirical research, how the variables were measured, and how the data was analyzed.

### 3.1 Research Design

For a period of 90 minutes, the participant works on a software development tasks of a real software project. The researcher observes the behavior of the individual while programming. Each 10 minutes, the participant completes a short questionnaire on a tablet device. That is, valence, arousal, and dominance are measured for 9 times per participant. The same holds for the self-assessment of the productivity.

Each participant faces a pre-task interview in which basic demographic data, information about the project, tasks, and the developer's skills are obtained. Descriptive data is collected on the role of the participant (either "professional" or "student"), the experience with the programming language, and experience with the task (low, medium, and high).

After the completion of the working period, the researcher conducts a post-task interview. The instructions given to each participant are available in the on-line Appendix [12] of this paper. We wrote the instructions for the Self-Assessment Manikin (SAM) questionnaire following the technical manual by Lang et al. [18].

The researcher is present during the entire development period to observe the behavior of the participant without interfering. During the post-task interview, the observer and the participant look at a generated graph of the participant's productivity. If there are noticeable changes in the data trends, especially if these changes are in conflict with the observer's notes and predictions, the participant is asked to explain what happened in that interval. Complete anonymity is ensured to the participants.

The context of this study is natural settings (i.e., the working environment). Participants are obtained from the students of Computer Science of the Free University of Bozen-Bolzano and local IT companies. There are no restrictions in gender, age, or nationality. Participation is voluntary and not rewarded.

The only required instrument is a suitable device that implements the SAM questionnaire and the productivity item. We designed a website that implements SAM, and it is optimized for tablet devices. Since SAM is a pictorial questionnaire, the effort required for the questionnaire session is thus reduced to 4 touches to the screen.

All steps of the experiment are automated.

### 3.2 Constructs and Measurements

The affective states dimensions - valence, arousal, dominance - describe differences in affective meanings among stimuli and are measured with the SAM pictorial questionnaire. The values of the affective state constructs range from 1 to 5. A value of 3 means "perfect balance" or "average" between the most negative (1) and the most positive value (5). For example, a value of 1 for the valence variable means "complete absence of attractiveness".

The task productivity is self-assessed by the participant, using a 5-point Likert item. The item is the sentence "My productivity is ..." The participant ends the sentence, choosing the proper ending in the set {very low, below average, average, above average, very high}.

Each participant's data is converted to the individual's Z-score for the set of construct measurements, using the formula in (1):

$$z_{score}(x_{pc}) = \frac{x_{pc} - \mu_{pc}}{\sigma_{pc}} \quad (1)$$

where $x_{pc}$ represents the measured participant's construct, $\mu_{pc}$ is the average value of all the participant's construct measurements, and $\sigma_{pc}$ is the standard deviation for the participant's construct measurements.

The measurements of each participant become dimensionless (strictly speaking, the unit is "number of standard deviations above or below the mean") and comparable, as they indicate how much the values spread. The range of the variables, while theoretically infinite, is practically the interval [-3, +3] due to the three-sigma rule [26].

### 3.3 Analysis Procedure

This study compares repeated measurements of individuals. The repeated measurements have a non-trivial impact on the analysis phase because 1) the data provided by each participant have dependencies among them, and 2) there might be time effects on the series of measurements per each participant. Thus, we have dependencies of the data at the participants' level and at the time level, grouped by the participant. Such dependencies present issues when employing Anova procedures, which are not designed for repeated measurements and multiple levels of dependency. Anova procedures are discouraged in favor of mixed-effects models, which are robust and specifically designed for repeated measurements and longitudinal data [14].

A linear mixed-effects model is a linear model that contains both fixed effects and random effects. The definition of a linear mixed-effects model given by Robinson [27] is given in (2):

$$y = X\beta + Zu + \varepsilon \quad (2)$$

where $y$ is a vector of observable random variables, $\beta$ is a vector of unknown parameters with fixed values (i.e., fixed effects), $u$ is a vector of random variables (i.e., random effects) with mean $E(u) = 0$ and variance-covariance matrix $var(u) = G$, $X$ and $Z$ are known matrices of regressors relating the observations $y$ to β and $u$, and $\varepsilon$ is a vector of independent and identically distributed random error terms with mean $E(\varepsilon) = 0$ and variance $var(\varepsilon) = 0$.

The estimation of the significance of the effects for mixed models is an open debate. A convenient way to express the significance of the parameters is to provide upper and lower bound p-values[1]. We implement the model using the open-source statistical software *R* and the *lme4.lmer* function for linear mixed-effects models.

## 4 Results

The designed data collection process was fully followed. No deviations occurred. The participants were fully committed.

### 4.1 Descriptive Statistics

We obtained eight participants, for a total of 72 measurements. The mean of the participants' age was 23.75 (standard deviation=3.29). Seven of them were male. Four participants were first year B.Sc. Computer Science students and four of them were professional software developers. The Computer Science students worked on course-related projects. The four professional software developers developed their work-related projects.

**Table 1.** Participants and Projects Details

| id | gender | age | role | project | task | p. lang. | p.lang exp. | task exp. |
|---|---|---|---|---|---|---|---|---|
| P1 | M | 25 | PRO | Data collection for hydrological defense | Module for data displaying | Java | HIG | HIG |
| P2 | M | 26 | PRO | Research Data Collection & Analysis | Script to analyze data | Python | LOW | HIG |
| P3 | M | 28 | PRO | Human Resources Manager for a School | Retrieval and display of DB data | Java | HIG | HIG |
| P4 | M | 28 | PRO | Metrics Analyzer | Retrieval and sending of metrics | C++ | HIG | HIG |
| P5 | F | 23 | STU | Music Editor | Conversion of music score to pictures | C++ | LOW | LOW |
| P6 | M | 20 | STU | Code Editor | Analysis of Cyclomatic Complexity | C++ | LOW | LOW |
| P7 | M | 20 | STU | CAD | Single-lined labels on objects | C++ | LOW | LOW |
| P8 | M | 20 | STU | SVG Image Editor | Multiple objects on a circle or ellipse | C++ | HIG | HIG |

---

[1] We advise to read the technical manual by Tremblay et al.: http://cran.r-project.org/web/packages/LMERConvenienceFunctions/

The characteristics of the participants are summarized in Table 1. We notice that the roles do not always correspond to the experience. The professional participant P2 reported a LOW experience with the programming language while the student participant P8 reported a HIGH experience in both the programming language and the task type. Table 1 also contains the characteristics of the projects and the implemented task. There is high variety of project types and tasks. Five participants programmed using C++ while two of them with Java and the remaining one with Python. The participants' projects were non-trivial, regardless of their role. For example, participant P1 (a professional developer) was maintaining a complex software system to collect and analyze data from different sensors installed on hydrological defenses (e.g., dams). Participant P5 (a student) was implementing pictorial exports of music scores in an open-source software for composing music.

We provide in Fig. 1, in Fig. 2, and in Fig. 3, the charts representing the changes of the self-assessed productivity over time, with respect to the valence, the arousal, and the dominance dimensions respectively.

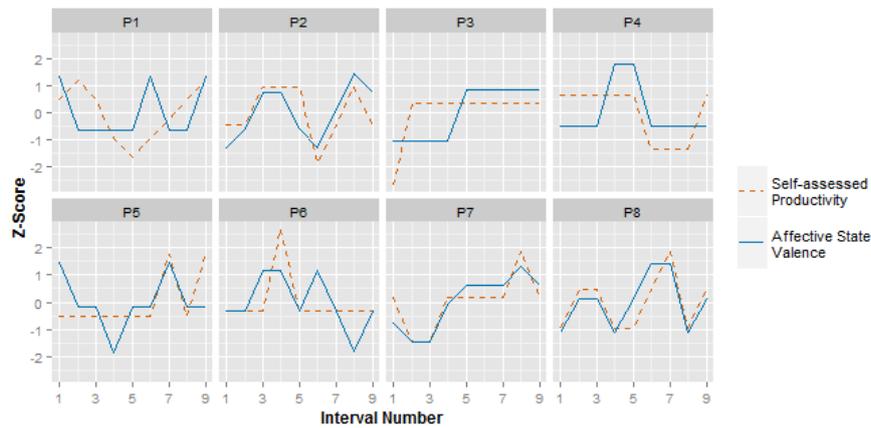

**Fig. 1.** Valence vs. Productivity over Time

As it can be seen in Fig. 1, there are cases in which the valence score provides strong predictions of the productivity (participants P2, P7, and P8). For the other participants, there are many matched intervals, e.g. P5 at interval 7, and P4 at intervals 4-7. Participant P1 is the only one for which the valence does not provide strong predictions. In few cases, the valence Z-score is more than a standard deviation apart from the productivity Z-score.

The arousal dimension in Fig. 2 looks less related to the productivity than the valence dimension. The behavior of the arousal line often deviates from the trend of the productivity line (e.g., all the points of participants P5 and P6). Nevertheless, there are intervals in which the arousal values are closely related to productivity, e.g., with participants P4 and P7.

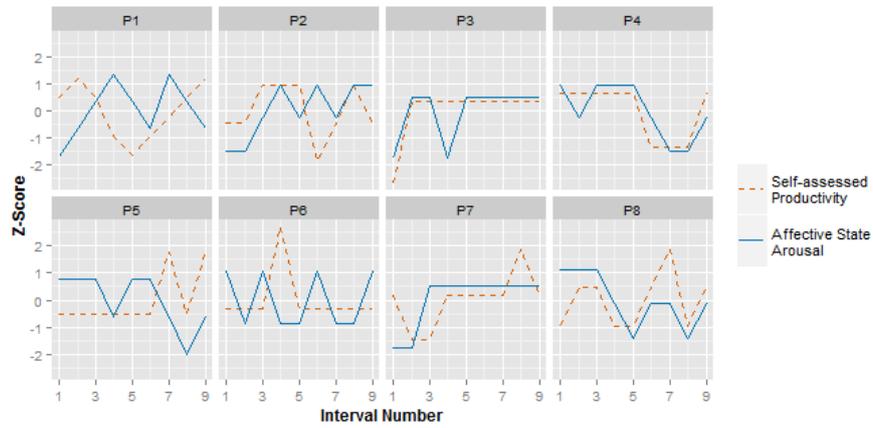

**Fig. 2.** Arousal vs. Productivity over Time

The dominance dimension in Fig. 3 looks more correlated to the productivity than the arousal dimension. Participants P1, P5, and P7 provided close trends. For the other cases, there are intervals in which the correlation looks closer and stronger. However, it becomes weaker for the remaining intervals (e.g., with P4). The only exception was with participant P6, where a clear correlation between dominance and productivity cannot be spotted.

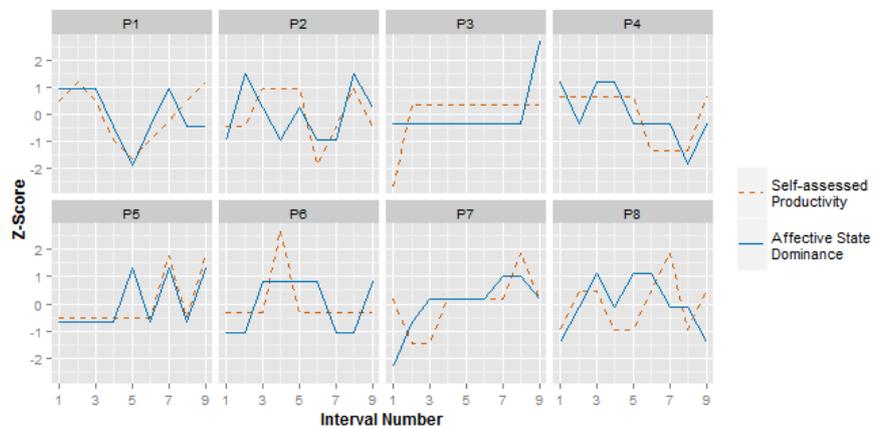

**Fig. 3.** Dominance vs. Productivity over Time

For all the participants, the Z-values of the variables show variations of about 2 units over time. That is, even for a working time of 90 minutes there are strong variations of both the affective states and the self-assessed productivity.

### 4.2 Hypotheses Testing

For the model construction, valence, arousal, dominance, and their interaction with time are modeled as fixed effects. The random effects are two: a scalar random effect for the participant grouping factor (i.e., each participant) and a random slope for the measurement time, indexed by each participant. In this way, the dependency of the measurements within the participants are taken into account: at the participant's level and at a time level. The final, full model[2] is given in (3) as a *lme4.lmer* formula.

productivity ~ (valence + arousal + dominance) ∗ time + (1 | participant) +
(0 + time | participant)     (3)

where *productivity* is the dependent variable; *valence*, *arousal*, *dominance*, and *time* are fixed effects; (1 | *participant*) is the scalar random effect for each participant, (0 + *time* | *participant*) is read as "no intercept and time by participant" and it is a random slope for the measurement time, grouped by each participant.

The full model in (3) significantly differs from the null model (4)

productivity ~ 1 + (1 | participant) + (0 + time | participant)     (4)

We checked for normality and homogeneity by visual inspections of a plot of the residuals against the fitted values, plus a Shapiro-Wilk test.

Table 2 provides the parameter estimation for the fixed effects (expressed in Z-scores), the significance of the parameter estimation, and the percentage of the deviance explained by each fixed effect. A single star (*) highlights the significant results (p-value less than 0.01). At a 0.01 significance level, valence and dominance are positively correlated with the self-assessed productivity of software developers.

**Table 2.** Parameter Estimation

| *Fixed Effect* | *Value* | *Sum Square* | *F-value* | *Upper p-value (64 d.f.)* | *Lower p-value (48 d.f.)* | *Deviance Explain. (%)* |
|---|---|---|---|---|---|---|
| valence | 0.10* | 7.86 | 19.10 | 0.000 | 0.000 | 12.39 |
| arousal | 0.07 | 0.00 | 0.00 | 0.950 | 0.950 | 0.00 |
| dominance | 0.48* | 7.44 | 18.07 | 0.000 | 0.000 | 11.71 |
| time | 0.00 | 0.11 | 0.26 | 0.614 | 0.615 | 0.17 |
| valence:time | 0.04 | 0.35 | 0.84 | 0.363 | 0.364 | 0.54 |
| arousal:time | -0.03 | 0.45 | 1.09 | 0.300 | 0.301 | 0.71 |
| dominance:time | -0.01 | 0.06 | 0.15 | 0.699 | 0.700 | 0.10 |

---

[2] Please note that if the intercept for the *time* parameter (i.e., (1+*time* | *participant*)) is not suppressed, the resulting model will be less valuable in terms of likelihood ratio tests (*anova* in R). Additionally, the value of the added random intercept would belong to the interval [-0.02, 0.03]. We thank an anonymous reviewer for the valuable feedback that let us improve this section.

The scalar random effects values for the participants belonged to the interval [-0.48, 0.33]; the random effects for the time were estimated to 0.

There is significant evidence to support the hypotheses H1 and H3. There is a positive correlation between two affective states dimensions (valence, dominance) and the self-assessed productivity of software developers.

Although we do not have evidence to support H2, regarding arousal, we will provide a possible explanation for this in the next section.

## 5 Discussion

In this section, we discuss the results and compare them with the related work. After the discussion of the results, we reflect on the limitations of the study.

### 5.1 Implications

The empirical results obtained in this study support the hypothesized positive correlation between the affective state dimensions of valence and dominance with the self-assessed productivity of software developers. No support was found for a positive correlation with the arousal affective state dimension and productivity. No evidence was found for a significant interaction between affective states and time.

The linear mixed-effects model provides an explanation power of 25.62% in terms of percentage of the deviance explained. Valence was estimated to 0.10 and dominance to 0.48, in terms of Z-scores. However, the percentage of the deviance explained by the two effects is almost the same: 12.39 for valence and 11.71 for dominance. In other words, high happiness with the task and the sensation of having adequate skills roughly have the same explanation power and provide almost the full explanation power of the model.

Regarding arousal, we suspect that the participants might have misunderstood its role in the questionnaire. All participants raised questions about the arousal dimension during the questionnaire explanations. A possible explanation of no significant interactions between the affective states dimensions and time is that each participant worked on different, independent projects. Also, the random effects related to time were estimated to 0, thus non-existing. It is worthy to report the full model with time as fixed and random effect because future experiments with a group of developers working on the same project will likely have significant interactions with time.

The results of this study are in line with the results of Khan et al. [16], where high valence resulted in the best performance on software debugging. However, the results are not in line with this study regarding the arousal dimension. The results of this study are also in line with those of Fisher and Noble [11], where positive affective states of different types of workers are found to be positively correlated with their productivity.

Although it is difficult to define productivity for software developers, all the participants had a clear idea about their productivity scores. None of them raised a question about how to self-rate their productivity. Nevertheless, in the post-task interviews, they were not able to explain how they defined their own productivity. The most common answer was related to the achievement of the expectation they set for the end of the 90 minutes of work. Again, this "expectation" was neither clearly definable nor quantifiable for them. Their approach was to express the sequent productivity value with respect to the previous one – as in "worse", "equal", or "better" than before.

The theoretical implication of this study is that the real-time affective states related to a software development task are positively correlated with the programmer's self-assessed productivity.

### 5.2 Limitations

In this section, we discuss the limitations of this study. We mitigated conclusion, internal, construct, and external validity threats while following the classification provided by Wohlin et al. [35].

*Conclusion validity* threats occur when the experimenters draw inaccurate inference from the data because of inadequate statistical tests. The employed linear mixed-effects models are robust to violations of Anova methods given by multiple dependencies of the data (see section 3.3). A threat lies in the limited number of participants (8) who worked for about 90 minutes each. However, the background and skills in the sample were balanced. Due to the peculiarity of the repeated measurements and the analysis method, all 72 measurements are valuable. It has been shown that repeated measures designs do not require more than seven measurements per individual [34]. We added two more measurements in order to be able to remove possible invalid data.

*Internal validity* threats are experimental issues that threaten the researcher's ability to draw inference from the data. Although the experiment was performed in natural settings, the fact the individuals were observed and the lack of knowledge about the experiment contents mitigated social threats to internal validity. A series of pilot studies with the measurement instrument showed that the minimum period to interrupt the participants was about 10 minutes if the case study was focused on a single task instead of longer periods of observations.

*Construct validity* refers to issues with the relation between theory and observation. A construct validity threat might come from the use of self-assessed productivity. In spite of the difficulty in using traditional software metrics (the project, the task, and the programming language were random for the researcher) and that measuring software productivity is still an open problem, self-assessed performance is commonly employed in Psychology studies [3, 11, 36] and it is consistent to objective measurements of performance [21]. We also carefully observed the participants during the programming task. Post-task interviews included questions on their choices for productivity levels, which resulted in remarkably honest and reliable answers, as expected.

*External validity* threats are issues related to improper inferences from the sample data to other persons, settings, and situations. Although half of the participants were students, it has been argued that students are the next generation of software professionals; they are remarkably close to the interested population if not even more updated on new technologies [17, 32]. Secondly, it can be questioned why we studied software developers working alone on their project. People working in group interact and trigger a complex, powerful network of affective states [2]. Thus, to better control the measurements, we chose individuals working alone. However, no participant was forced to limit social connections while working, and the experiment took place in natural settings.

## 6      Conclusions

For more than thirty years, it has been claimed that software developers are essential when considering how to improve the productivity of development process and the quality of delivered products. However, little research has been done on how human aspects of developers have an impact on software development activities. We echo a call on employing psychometrics in Software Engineering research by studying how the affective states of software developers - emotions, moods, and feelings - have an impact on their programming tasks.

This paper reports a repeated measures research on the correlation of affective states of software developers and their self-assessed productivity. We observed eight developers working on their individual projects. Their affective states and their self-assessed productivity were measured on intervals of ten minutes. A linear mixed-effects model was proposed in order to estimate the value of the correlation of the affective states of valence, arousal, and dominance, and the productivity of developers. The model was able to express about the 25% of the deviance of the self-assessed productivity. Valence and dominance, or the attractiveness perceived towards the development task and the perception of possessing adequate skills, were able to provide almost the whole explanation power of the model.

The understanding provided by this study is an important part of basic science in Software Engineering rather than leading to direct, applicable results: among Khan et al. [16] and Shaw [31], this is one of the first studies examining the role of affective states of software developers. We are providing basic theoretical building blocks on researching the human side of software construction. This work performs empirical validation of psychometrics and related measurement instruments in Software Engineering research. It proposes the employment of linear mixed-effects models, which have been proven to be effective in repeated measures designs instead of Anova. It is also stressed out that Anova should be avoided in such cases.

Are happy developers more productive? The empirical results in this study indicate towards a "*Yes, they are*" answer. However, a definite answer will be provided by future research characterized by the use of multidisciplinary theories, validated measurement instruments, and analysis tools as exemplified in this paper.

Experiments with a larger number of participants performing the same programming task will allow the use of traditional software productivity metrics and provide further explanations. Mood induction techniques should be employed to study causality effects of affective states on the productivity of software developers. Additionally, future studies on software teams with affective states measurements are required in order to understand the dynamics of affective states and the creative performance of software developers.

Software developers are unique human beings. By studying how they perceive the development life-cycle and how cognitive activities affect their performance, we will open up a different perspective and understanding of the creative activity of software development.

**Acknowledgment** We would like to thank the participants of the experiment. We would also like to acknowledge Elena Borgogno, Ilenia Fronza, Cigdem Gencel, Nattakarn Phaphoom, and Juha Rikkilä for their kind help and support during this study. We are grateful for the insightful comments offered by three anonymous reviewers, who helped us to improve the paper.